\documentstyle[aps,epsf,twocolumn,prl]{revtex}

\begin{document}
\draft

\twocolumn[\hsize\textwidth\columnwidth\hsize\csname @twocolumnfalse\endcsname
%%%%%%%%%%%%%%%%%%%%%%%%%%%%%%
\title{Solution of the Quantum Sherrington-Kirkpatrick Model}

\author{D. R.  Grempel$^1$ and M. J. Rozenberg$^2$}

\address{$^1$CEA/D\'epartement de Recherche Fondamentale sur
la Mati\`ere Condens\'ee \\
SPSMS, CENG, 17, rue des Martyrs, 38054 Grenoble
Cedex 9, France \\ $^2$Institut Laue--Langevin, BP156, 38042 Grenoble, France
}

\date{\today}
\maketitle
\widetext
\begin{abstract}
\noindent
%        1         2         3         4         5         6         7         8
%2345678901234567890123456789012345678901234567890123456789012345678901234567890
We solve the $S=1/2$ infinite-range random Heisenberg
Hamiltonian in the paramagnetic phase
using quantum Monte Carlo and analytical techniques. 
We find that the spin-glass 
susceptibility diverges at a finite temperature $T_g$
which demonstrates the existence of a low-temperature ordered phase. 
Quantum fluctuations reduce the critical temperature and the effective Curie 
constant with respect to their classical values. 
They also give rise to a redistribution of spectral
weight in the dynamic structure factor in the paramagnetic phase. 
As the temperature decreases the spectrum of magnetic excitations
gradually splits into quasi-elastic and inelastic contributions 
whose weights scale
as $S^2$ and $S$ at low temperature.
\end{abstract}

\pacs{75.10.Jm, 75.40.Gb, 75.10.Nr}

]

\narrowtext

It is well known that the 
combined effects of randomness and frustration may lead 
to spin glass behavior in classical disordered magnets at low temperature. 
The most widely studied spin-glass Hamiltonian 
is the Sherrington-Kirkpatrick model\cite{sk}. 
However, only a small fraction of the vast amount of 
work\cite{reviews} devoted to this model directly addresses the role of 
quantum fluctuations. In a notable 
early paper Bray and Moore \cite{bm} first formulated
the theory of the quantum 
Sherrington-Kirkpatrick model and showed that it 
reduces exactly to an effective single site problem in
imaginary time. 
Using a variational approach, these authors argued that a spin
glass ordered phase occurs  
below a finite critical temperature for all values of $S$. 
Much more recently, Sachdev and Ye \cite{subir} 
discussed a generalized spin-glass Hamiltonian in which the spin components 
become the generators of 
the group SU($M$) and the states span a 
representation of the group labeled by an 
integer $n_b$ ($M=2$ and $n_b=2S$ for physical spin-$S$ operators). 
The model 
can be solved exactly when $M$ and $n_b \rightarrow \infty$ 
with $\kappa \equiv n_b/M$ finite. In this limit,
Sachdev and Ye argued that the ground state is either
a spin glass for large values of $\kappa$ (that plays the role of $S$), 
or a spin fluid below a critical value $\kappa_c$. 
In the spin fluid phase the local dynamic 
susceptibility exhibits unconventional 
behavior $\sim \ln({1 \over \omega})$ at $T=0$.

In view of these results, the nature of the ground state of the quantum 
Sherrington-Kirkpatrick model remained an open problem, the main question
being whether quantum fluctuations for low enough
$S$ may prevent the instability towards spin-glass order present
in the classical case \cite{fluc}.

In this paper we answer this question by means of an exact 
numerical solution of the $S=1/2$ model 
using a quantum Monte Carlo technique, and obtain analytical expressions for
the asymptotic forms of the imaginary part of the
local dynamic spin 
susceptibility $\chi''(\omega)$ in the limits $T \to 0$ and
$T \to \infty$. We find that the paramagnetic solution is 
unstable towards spin glass order at a finite temperature $T_g$. 
The transition temperature and the effective Curie constant 
are reduced with respect to 
their classical values by quantum effects. 
The analysis of the paramagnetic dynamic correlation function 
shows that the 
spectrum of magnetic excitations splits at low temperatures into two 
well defined contributions. 
The first one is similar to the 
low-frequency dynamic response function of   
classical fluctuating paramagnets\cite{forster}. 
The second one is an inelastic contribution
at higher frequencies, characterized by a large energy scale. 
At high $T$,
all the spectral weight $S(S+1)$ is concentrated in the
quasi-elastic feature. With decreasing temperature part
of the weight is transferred from low to 
high energies until, when $T \to 0$, the intensities
of the quasi-elastic and inelastic
parts reach the asymptotic values $S^2$ and $S$, respectively.  

The Sherrington-Kirkpatrick Hamiltonian is,
\begin{eqnarray}
H = -{1\over \sqrt{N}} \sum_{i<j} J_{ij}  \  \vec{S}_{i} \cdot \vec{S}_{j},  
\label{hamil}
\end{eqnarray}
where $\vec{S}_i$ is a three dimensional 
spin-1/2 operator at $i-$th site of a lattice of size $N$.
The exchange interactions $J_{ij}$ are independent 
random variables with a gaussian 
distribution with zero 
mean and variance $J=\langle J_{ij}^2\rangle^{1/2}$.

In the limit $N\rightarrow \infty$,
the  free-energy per spin 
${\cal F}$ has been derived 
by Bray and Moore\cite{bm} using  the replica method 
to do the average over
the quenched disorder and a 
Hubbard-Stratonovich transformation to decouple the different sites, 

\begin{eqnarray}
\nonumber
\beta {\cal F}=\min_{Q(\tau)}
\left\{
\frac{3J^2}{4}\int_{0}^{\beta}\int_{0}^{\beta}d\tau d\tau' 
Q^{2}(\tau-\tau') 
\right.
\end{eqnarray}
\begin{eqnarray}
\label{free-energy}
\left.
 -\ln {\rm Tr T}\exp\left[{J^2\over 2}\int_{0}^{\beta}
\int_{0}^{\beta}d\tau d\tau'Q(\tau-\tau')\vec{S}(\tau)
\cdot\vec{S}(\tau')\right]
\right\}.
\end{eqnarray}
Here, T is the time-ordering operator along the imaginary-time axis, 
$0\le \tau \le \beta$, and the trace is 
taken over the eigenstates of the spin-$1/2$ 
operator. The local dynamic correlation 
function in imaginary-time $Q(\tau)$, 
is determined by functional minimization of (\ref{free-energy}), 

\begin{eqnarray}
\label{selfconsist}
Q(\tau) = \frac{1}{3} \langle {\rm T}
\vec{S}(\tau)  \cdot \vec{S}(0)\rangle,
\end{eqnarray}
where the thermal average is taken with respect to the 
probability associated to the local partition function, 

\begin{eqnarray}
\label{zloc}
Z_{loc}={\rm Tr T} \exp\!\left[
\frac{J^2}{2}\!\int_{0}^{\beta}\!\int_{0}^{\beta}\!
d\tau d\tau'Q(\tau-\tau')\vec{S}(\tau)\!\cdot\!\vec{S}(\tau')\right].
\end{eqnarray}

The spin-glass transition temperature may be calculated from local 
quantities. It results from the instability criterion\cite{bm}  
$J\chi_{loc}=1$, with the static local 
susceptibility $\chi_{loc}= \int_0^\beta d\tau Q(\tau)$.

In order to set up a numerical scheme for the solution 
of the model we find it necessary 
to perform an additional
Hubbard-Stratonovich transformation 
and rewrite $Z_{loc}$ as

\begin{eqnarray}
Z_{loc}= && \int {\cal D}\vec\eta \exp\!
\left[-{1\over 2}\!\int_0^\beta\!\int_0^\beta\!d\tau d\tau'
Q^{-1}(\tau,\tau')\vec{\eta}(\tau)\!\cdot\!\vec{\eta}(\tau')\right]
\nonumber\\
&& \times {\rm Tr T}\exp
\left[
\int_0^\beta\!d\tau J\vec\eta(\tau)\!\cdot\!\vec{S}(\tau)
\right].
\label{action}
\end{eqnarray}
$Z_{loc}$ is the average partition function 
of a spin in an effective ``time''-dependent 
random magnetic field $J\vec{\eta}(\tau)$ distributed with a gaussian
probability.  
This formulation is well suited 
to implement a quantum Monte Carlo algorithm. 
The imaginary time axis is discretized into $L$ time slices and the 
time-ordered exponential under the trace in Eq. \ref{action} 
is written 
as the  product of $L$ matrices of $2\times 2$ using Trotter's 
formula. 
We performed calculations 
for $\beta \le 50$ and $L\le 128$ (keeping 
$J\Delta \tau = J\beta/L\le 0.5$). 
There are two important technical remarks: firstly, it is crucial that 
each trajectory ${\vec \eta(\tau) }$ 
and its time-reversed partner  
${\vec \eta(\beta-\tau)}$ be considered simultaneously
in order to obtain a real probability measure\cite{we}. 
Secondly, the sorting procedure is formulated 
in the frequency domain: the phase space for the simulation 
consist of all the realizations
of ${\vec \eta (\omega_n)}$ with the integer $n$ labelling the 
bosonic Matsubara frequencies. This new set of variables presents
the advantage of
being much less 
correlated than the original one. An elemental Monte Carlo move
is thus to propose a 
change in the complex field $\eta_i (\omega_n)$ for given
$i =x,y,z $ and $n$;  a full update of the system is completed
after elemental moves 
have been attempted for all directions and frequencies.
The numerical procedure is as follows: i) an initial $Q(\omega_n)$ is used 
as input in (\ref{action}). ii) the spin-spin
correlation function is obtained using Monte Carlo. iii) a new
$Q(\omega_n)$ is calculated from the self-consistency condition
(\ref{selfconsist}) and used as a new input in step i). 
This procedure is repeated until convergence is attained which 
typically occurs after 5 iterations.  
The main source of error in the results comes from the statistical noise 
due to the random sampling. 
The efficiency of our code allowed us to sensibly reduce
it by performing $10^{5}$ full updates in the last iteration. 
It is remarkable that we faced no ``sign 
problem'' even down to the lowest temperature considered, $T= 0.02J$. 
\begin{figure}
\epsfxsize=3.5in
\epsffile{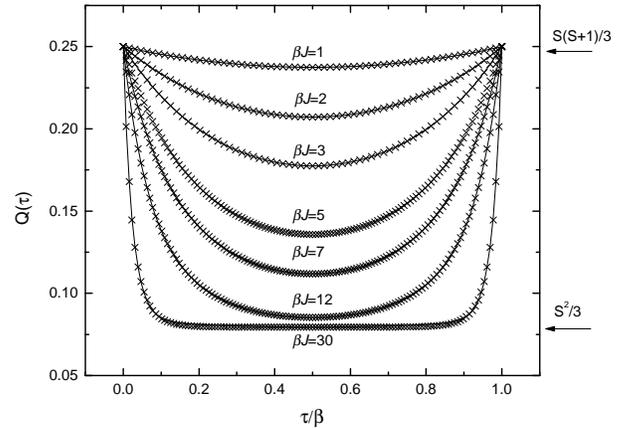}
\caption{$Q(\tau)$ as function of $\tau /\beta$ for several values of
$\beta J$. The crosses correspond to QMC data. The error bars are well
below the size of the crosses. Solid lines are fits obtained with the
function $Q_{par}(\tau)$ described in the text.
}
\label{fig-1}
\end{figure}
In Fig. \ref{fig-1} we show the 
correlation function obtained with this method
which exhibits the following qualitative features.
For $T>>J$, $Q(\tau)$ is nearly constant and 
close to $S(S+1)/3=1/4$, its classical limit. 
In contrast, for $T<<J$, $Q(\tau)$ rapidly decreases at 
both ends of the interval $[0,\beta]$ and then varies slowly remaining
near the value $S^2/3$.
The behavior in the intermediate 
temperature range is a smooth interpolation
between the two extreme cases. 
We obtain $\chi_{loc}$
%the local static susceptibility 
by numerically integrating $Q(\tau)$ and find that it crosses over between 
two limiting 
forms, $\chi_{loc} \approx S(S+1)/(3T)$ for $T>>J$ and $\chi_{loc} \approx 
S^2/(3T)$ for $T<<J$, implied by the asymptotic 
behavior of $Q(\tau)$ described above. The susceptibility 
thus obeys a Curie law at both ends of the temperature range with the 
low $T$ effective Curie constant reduced by a factor of 3 with respect to
its classical value. 
Note that the low-temperature behavior of
$\chi_{loc}(T)$ differs considerably from the 
prediction of the large-$M$ model\cite{subir}  
which is $\chi_{loc}(T)\sim \ln(1/T)$ for 
small $\kappa$.
On the other hand, the full temperature
dependence of $\chi_{loc}$ is remarkably close to that 
predicted by the variational approach\cite{bm}. 
This is easily understood at high and low temperatures where the
variational ansatz of Bray 
and Moore\cite{bm}, $Q(\tau) = const$, represents well the actual numerical 
solution (cf. Fig.\ref{fig-1}). However, the close
agreement at intermediate temperatures is rather unexpected.
The criterion for the appearance of
spin glass order, $J\chi_{loc}=1$, is 
fulfilled at $T_g \approx 0.142 J$ 
that compared to the classical value $T_g=0.25 J$ shows
the importance of quantum fluctuations in this system.

We now turn to the discussion of the frequency dependence of the 
local susceptibility. 
This involves extracting real-frequency information
from results obtained on the imaginary axis,
which is in general a rather difficult task. 
However, it will be seen below that in the 
present case the dissipative part of 
the local response $\chi''_{loc}(\omega)$ may be found analytically 
at low and 
high temperature. This allows us to construct an interpolating
function that accurately
describes our numerical data throughout the 
whole temperature range, enabling us to extract
physical information from the imaginary-time Monte Carlo results.

The dynamics of the system for $T >>J$ is
on general grounds expected to be 
controlled by a single relaxation rate ${\omega_L=\cal O}(J)$. 
This assumption implies

\begin{eqnarray}
\label{chi"}
\frac{\chi''_{loc}(\omega)}{\pi \omega}=
\chi_{loc}\frac{1}{\omega_{L}} F(\frac{\omega}{\omega_{L}}),
\end{eqnarray}
where the relaxation function $F$ is constrained to be normalized to one 
and have
a finite second moment\cite{forster}. The simplest function  
fulfilling these conditions is a gaussian. 
With this ansatz, the response is completely determined by the sum-rule 

\begin{eqnarray}
\label{sum-rule}
\int_{-\infty}^{\infty}\frac{d\omega}{\pi}\omega 
\chi''_{loc}(\omega)=2J^{2}\int_{0}^{\beta}Q^{2}(\tau), 
\end{eqnarray}
that is derived\cite{we} using the generic
$f$-sum rule for spin systems and (\ref{free-energy}). 
%for the free energy of the model. 
It then follows from (\ref{chi"}) and (\ref{sum-rule})
that 
\begin{eqnarray}
\label{quasi-elastic}
\frac{\chi''_{loc}(\omega)}{\pi \omega}
=\frac{\beta S(S+1)}{3}
\left[1- S(S+1){(\beta J)^2 \over 18}\right]
\frac{e^{-\frac{1}{2} 
(\frac{\omega}{\omega_L})^2}}{
{\sqrt{2\pi \omega_L^2}}},
\end{eqnarray}
with $\omega^{2}_{L} = 2J^{2}S(S+1)/3[1- S(S+1)(\beta J)^2/18]$ 
for $T >> J$. 
It can be shown\cite{we}
that Eq. \ref{quasi-elastic} reproduces the first 
few orders of the high-temperature expansion of $Q(\tau)$. 
This expression is thus correct for $T>>J$.

With decreasing temperature, 
the assumption of a single relaxation rate breaks down.
At $T<<J$ the existence of two different characteristic times suggested 
by the numerical data of Fig. \ref{fig-1} must be reflected 
 in the emergence of well separated
energy scales in the frequency dependent dynamic response.
Indeed, this behavior follows from an approximate analytical solution of 
Eqs. \ref{selfconsist} and \ref{action} 
that can be shown to be exact in the $T \to 0$ limit.
{}From Eq. \ref{action} the problem can be thought of as that of a single spin
in a fluctuating effective magnetic field $J\vec \eta(\tau)$. 
We will see below that at low $T$
this effective field is dominated by its
$\omega_n=0$ component. 
Therefore, it is convenient to split the effective field
into a constant part $\vec{h}_{0}=J\vec{\eta}(\omega_n=0)/\sqrt{\beta}$
and a small $\tau$-dependent part with 
$|\delta\vec{h}(\tau)| << |\vec{h}_0|$. We thus write
(\ref{selfconsist}) as 
\begin{equation}
Q(\tau)={1 \over 3}\left[ \langle Q_{L}(\tau,\vec{h}_{0})
\rangle_{\vec{h}_0} + 2 
\langle Q_{T}(\tau,\vec{h}_{0})\rangle_{\vec{h}_0}\right],
\label{decomposition}
\end{equation}
where $Q_{L}$ and $Q_{T}$ are, respectively, the 
longitudinal and transverse response functions in an applied 
field $\vec{h}_0$ having formally integrated out 
the fields $\delta{\vec{h}}(\tau)$.
The angular brackets denote the 
average with respect to the isotropic distribution 
$P(|\vec{h}_0|)$. 
For a given $\vec{h}_0$, the imaginary part of the transverse response
function $\chi''_{T}(\omega)$ has a peak at 
$\omega=|\vec{h}_0|$ whose width $\Gamma$ is
proportional to the square of 
the amplitude of the fluctuating field at the 
resonance frequency. 
A simple estimate\cite{note2} shows that for $T<<J$, 
$P(|\vec{h}_0|)$ is maximum at $|\vec{h}_0|\equiv \omega_T 
\approx J^{2}S\chi_{loc}$, 
which is large at low temperature. 
Assuming for the moment that
$|\delta\vec{h}(\tau)| << |\vec{h}_0|$ ({\it i.e.}, setting $\Gamma=0$) 
one can perform the average over $P(|\vec{h_0}|)$ and obtain
an approximation for the dissipative part of 
$\langle Q_{T}(\tau,\vec{h}_{0})\rangle_{\vec{h}_0}$,

\begin{eqnarray}
\nonumber
\frac{\chi''_{T}(\omega)}{\pi \omega}=
\frac{S}{2}
\left[\frac{\beta}{2 \omega_T}\right]^{3/2}
\frac{\omega\tanh(\beta \omega/2)}{\sqrt{2\pi}(1+\beta \omega_T/2)}
\end{eqnarray}
\begin{eqnarray}
\label{chi"tr}
\times
\left\{\exp\left[-\frac{\beta}{4 \omega_T}(\omega-\omega_T)^2\right]
+\exp\left[-\frac{\beta}{4 \omega_T}(\omega+\omega_T)^2\right]
\right\}.
\end{eqnarray}

Using this equation and the fluctuation-dissipation theorem we estimate 
$\Gamma/\omega_T={\cal O}(T/J)$, showing that the assumption leading
to (\ref{chi"tr}) is indeed correct. Therefore, 
the expression above is asymptotically exact for $T\to 0$.

The high-frequency scale $\omega_T\sim J^2/T$ where $\chi''_{T}(\omega)$
is sharply peaked, is associated to
the initial rapid decrease in $Q(\tau)$ 
observed in our low-$T$ simulations.
The remaining contribution to the dynamic response function 
$\chi''_L(\omega)$ comes from the relaxation of the longitudinal 
magnetization. As the amplitude
of the fluctuating field $|\delta{\vec{h}(\tau)}| << |\vec{h}_0|$
we expect this process to be slow.
Its frequency $\omega_L$ may be found irrespectively of 
the detailed form of 
$\chi''_L(\omega)$ using Eqs. \ref{sum-rule}, \ref{decomposition}
and \ref{chi"tr} which yield $\omega_L\propto T^2$ \cite{we}. This
small energy scale is associated to the slowly varying part 
of $Q(\tau)$
observed at low temperature. Using again the ansatz (\ref{chi"})
for $\chi''_L(\omega)$, 
the sum-rule (\ref{sum-rule}) leads to an 
expression similar to (\ref{quasi-elastic}), except that the prefactor of 
the exponential is now simply 
$\beta S^2/3$. 
Notice that the dynamics of magnetic fluctuations 
that emerges from the above arguments
bears no resemblance to that of the spin-fluid state of the large-$M$ 
model.

Comparison between the low- and high-temperature results
implies that when $T$ decreases, a fraction of the 
spectral weight of the quasi-elastic peak  
at small-$\omega$ is transferred to the
high energy excitations described by Eq. \ref{chi"tr}. 
This redistribution of intensity, closely related to   
the reduction of the effective Curie 
constant discussed above, is a distinctive
quantum effect: the strength of the inelastic feature relative to that of 
the quasi-elastic peak is 
${\cal O}(1/S)$ and vanishes in the large-$S$ limit.
The analytical results just discussed 
suggest a parametrization of the spin-spin
correlation function $Q(\tau)$ that contains the exact
asymptotic forms at both high and low $T$
and smoothly interpolates between them.
The proposed interpolation function is 
defined as $Q_{par}(\tau) = S(S+1)/ 3[ p
\Phi_{L}(\tau) + (1 -p ) \Phi_{T}(\tau)]$ where,

\begin{eqnarray}
\nonumber
\Phi_{L}(\tau)= e^{ -\frac{1}{2} \left(\frac{\beta \omega_L}{2}
\right)^2
\left[ 1 - \left( 1 - \frac{2 \tau}{\beta}\right)^2 \right]}
\end{eqnarray}
\begin{eqnarray}
\Phi_{T}(\tau)= \frac{1+ \beta \omega_T/2 \left(
1 - 2 \tau /\beta\right)^2}{1 + \beta \omega_T / 2}
e^{ -\frac{\beta \omega_T}{4}
\left[ 1 - \left( 1 - \frac{2 \tau}{\beta}\right)^2 \right]},
\end{eqnarray}
where $\Phi_{L}$ and $\Phi_{T}$ are the imaginary-time equivalents of
Eqs. \ref{quasi-elastic} and \ref{chi"tr} 
normalized such that $\Phi_{L,T}(0)=1$, and
$\omega_L$ and $\omega_T$ are now parameters
corresponding to the 
width of the central peak and the
characteristic scale of the high-energy excitations, respectively.
The third parameter, $p$, controls the transfer of spectral weight
between the two components of the magnetic response. At high $T$
there is a single energy scale and only 
quasi-elastic intensity is present as
$p \to 1$, while at low $T$ inelastic intensity appears and 
$p \to S/(S+1)$ its lower bound. 

\begin{figure}
\epsfxsize=3.5in
\epsffile{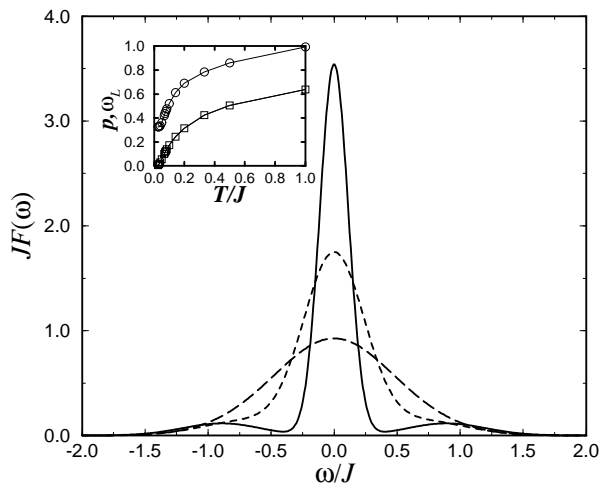}
\caption{The relaxation function $F(\omega) = \chi''(\omega) / 
(\pi \chi_{loc} \omega)$ for $\beta J$ = 3 (long dashed), 7 (dashed) and
14 (solid). Inset: $T-$dependence of the parameters $p$ (circles)
and $\omega_L$ (squares)
that have the limiting behavior predicted by the theory.
}
\label{fig-2}
\end{figure}

Using this expression we obtain highly accurate fits
of our numerical results as demonstrated
in Fig. \ref{fig-1}. While $Q_{par}(\tau)$ contains
the correct high and low $T$ limits by construction, 
it is remarkable that the quality
of the agreement remains excellent through {\em all} the temperature range. 
This gives us confidence that the essential physics 
is indeed captured by our parametrization of $Q(\tau)$.
It is now interesting to go back to real frequency and
plot $\chi''_{loc}(\omega)/\omega$ which is an
experimentally accessible quantity.
The results shown in Fig. \ref{fig-2} 
illustrate the gradual emergence of the high-frequency
features in the dynamic response. 

Our results are strictly valid above $T_g$, since below this temperature
the paramagnetic state is unstable \cite{note3}.
Nevertheless, the study of the solution in the whole $T-$range
is justified as it should be kept in mind that
the precise value of $T_g$ depends on the details of the
model Hamiltonian. For instance, lowering the dimensionality will
enhance the role of fluctuations which in turn are expected to reduce the
value of the transition temperature. 
Therefore, the qualitative properties of
our paramagnetic solution 
may be relevant for 
real quantum spin glasses in their disordered phase.
In this context two predictions that 
emerge from our work may provide useful
insight for the analysis of experiments on quantum frustrated
systems \cite{exp} provided that $T_g$ is small enough. 
i) Measurements of
the magnetic susceptibility as the transition is approached 
from above may indicate an anomalously low value of the 
effective Curie constant. ii) For $T \ge T_g$
a fraction of the 
spectral weight may be spread over a wide energy range and be
difficult to distinguish from background noise. 
This fact, combined with the presence 
of a very strong and narrow central peak, may result in 
an apparent loss of spectral weight
in neutron scattering experiments.

Many interesting questions remain to be addressed, 
for instance, the
effect of coupling the spin-system to an electronic band. 
For a bandwidth larger than
$J$ one may expect that $T_g \to 0$ leading to the interesting
physics of systems near quantum critical points \cite{sr}.


\begin{references}
\bibitem{note2} 
This estimate is be obtained by evaluating the probability
associated with Eq.\ref{action}
for a $\tau$-independent field.
\bibitem{note3} 
The same remark applies to
the spin fluid phase of the large $M$ Hamiltonian which 
cannot be the true ground state. 
The behavior $\chi_{loc} \sim \ln (1/T)$ 
implies that the criterion for a spin glass
transition $1= J\chi_{loc}$ is satisfied at a finite $T$.
\end{references}
\end{document}